\documentclass[preprint,10pt]{elsarticle}
\usepackage{graphicx}
\usepackage{amsmath}
\usepackage{amssymb} 
\usepackage{xurl}
\usepackage{lineno,hyperref}
\modulolinenumbers[5]
\usepackage{hyperref}

\DeclareMathOperator{\arccot}{arccot}

\journal{Annals of Physics}

\begin{document}

\begin{frontmatter}

\title{Elementary atoms in spaces of constant curvature by the Nikiforov-Uvarov method}

\author{Abdaljalel E. Alizzi}
\ead{abdaljalel90@gmail.com}
\address{Budker Institute of Nuclear Physics and Novosibirsk State University, Novosibirsk 630 090, Russia, and Al Furat University, Deir-ez-Zor, Syrian Arab Republic}

\author{Alina E. Sagaydak}
\ead{a.sagaidak@g.nsu.ru}
\address{Novosibirsk State University, Novosibirsk 630 090, Russia}

\author{Zurab K. Silagadze}
\ead{Z.K.Silagadze@inp.nsk.su}
\address{Budker Institute of Nuclear Physics and Novosibirsk State University, Novosibirsk 630 090, Russia}

\end{frontmatter}

\begin{abstract}
The Nikiforov-Uvarov method is a simple, yet elegant and powerful method for solving second-order differential equations of generalized hypergeometric type. In the past, it has been used to solve many problems in quantum mechanics and elsewhere. We apply this method to the classical problem of hydrogen-like atoms in spaces of constant curvature. Both the spectra of these atoms and their wave functions, including normalization, are easily obtained.
\end{abstract}

\maketitle

\section{introduction}
Before Lobachevsky and Bolyai, the question of whether there is another geometry besides Euclidean had not even arisen. Therefore, one might expect that the discovery of non-Euclidean geometry would have had the effect of an exploding bomb in scientific circles. But this is not at all the case. The acceptance of non-Euclidean geometry did not come easy, as is eloquently demonstrated by the letter of the Russian literary and social critic Nikolai Chernyshevsky to his sons from exile, in which he wrote that all of Kazan laughed at Lobachevsky \cite{Gindikin_1988}. Not a single serious scientist paid attention to Lobachevsky's publications, either in Russia, France or Germany. With one, but important, exception. Gauss read his short book in German and was so impressed that he began to study Russian, perhaps in order to read Lobachevsky's earlier publications in the Kazan University journal \cite{Prasolov_2015}. He succeeded in getting Lobachevsky elected as a corresponding member of the Royal Scientific Society of G\"{o}ttingen. However, despite Gauss's support, Lobachevsky died without having achieved recognition of his ideas \cite{Boltianski_2002}.

The situation began to change in the second half of the 19th century with the work of Beltrami, Cayley, Klein and Poincar\'{e}. Although the work of Lobachevsky and Bolyai was little understood in their lifetimes, they ultimately helped revolutionize almost every idea about geometry \cite{Gray_2007}. There is currently no clear-cut answer to the question of what geometry is because ``the meaning of the word {\it geometry} changes with time and with the speaker" \cite{Chern_1979}.

Both Lobachevsky and Bolyai proposed an analogue of Newton's force law for the hyperbolic space $\mathbb{H}^3$, and Bolyai even put forward a proposal to study the motion of planets around the Sun in this non-Euclidean hyperbolic 3-space \cite{Dombrovski_1991,Shchepetilov_2005}. Some research in the context of such a generalized Kepler problem followed from authors such as Schering, Lipschitz, Killing and Liebman, and these works can be considered as precursors of the special and general theories of relativity. However, after the rise of these pillars of modern physics, these works were almost completely forgotten for many years \cite{Shchepetilov_2005}. They are reviewed in \cite{Dombrovski_1991}.

A revival of interest in the Coulomb/Kepler problem in spaces of constant curvature occurred after Schr\"{o}dinger demonstrated the power of his newly discovered factorization method on the quantum mechanical problem of the hydrogen atom in the three-dimensional sphere $\mathbb{S}^3$ \cite{Schrodinger_1940}. Since then, a number of authors have studied this interesting question from both classical and quantum points of view \cite{Dombrovski_1991,Shchepetilov_2006,Ovsiyuk_2025,Borisov_2004,Nieto_1999,Redkov:2011,Carinena_2012,Redkov_2012,Ozfidan_2025} (In order not to excessively increase the bibliography of this short note, we will list only some representative works in which many other references can be found).

As is well known, the spectrum of the hydrogen atom in flat space is degenerate due to ``hidden" dynamical symmetry \cite{Maclay_2020}. Schr\"{o}dinger's results \cite{Schrodinger_1940} for $\mathbb{S}^3$, and then Infeld and Schild's results \cite{Infeld_1945} for $\mathbb{H}^3$ showed that degeneracy is also present in spaces of constant curvature. After Higgs' influential paper \cite{Higgs_1979}, this dynamical symmetry of the hydrogen atom in spaces of constant curvature attracted considerable attention. Usually dynamical symmetries are described mathematically in terms of Lie algebras.  It turned out that in spaces of constant curvature the most natural algebraic structures describing dynamical symmetries are quadratic algebras \cite{Granovskii_1992}, introduced by Sklyanin in 1983 \cite{Sklyanin_1983}.

Even the quantum mechanics of a free particle in spaces of constant curvature is more complicated than in flat space, since the canonical momenta do not coincide with the Noether momenta, and the latter do not commute \cite{Carinena_2011}. On the other hand, spaces of constant curvature provide the simplest curved background against which to study theoretical problems and questions related to quantum mechanics in curved spaces, and therefore such studies are of considerable theoretical interest. However, they are not only of academic interest. Effective curvature of space arises in a number of real physical situations. For example, in an interesting, though somewhat esoteric approach \cite{With_2023}, the melting of a crystal can be described as a transition to a space with constant negative curvature, for which the curvature is proportional to the density of disclinations in real physical space \cite{Novikov_1984}. The study of the motion of a quantum particle on a two-dimensional surface in condensed matter physics, as well as the study of quantum dots, has also led to the use of models based on quantum mechanics in spaces of constant curvature \cite{Carinena_2011}.
There are many other examples in the literature of the use of spaces of constant curvature in various fields: from the physics of atoms and nanotubes to the chiral and deconfinement phase transition in the Nambu-Jona-Lasinio model, symmetries of the $W$ algebra of string theory and quasi-exactly solvable models, as well as superintegrability in the framework of supersymmetry (see, for example, \cite{Kirchbach_2008,Quense_2023,Carinena_2021}).

About hydrogen atom in $\mathbb{S}^3$, Schr\"{o}dinger declared \cite{Schrodinger_1940} that he found the problem ``difficult to tackle in any other way" different from his factorization method.  But a year later Stevenson showed \cite{Stevenson_1941} that the spectrum and wave function could be obtained without too much difficulty by the usual methods of solving differential equations. An even simpler solution for the hydrogen-like atom in $\mathbb{S}^3$ and $\mathbb{H}^3$ can be obtained using the Nikiforov-Uvarov method \cite{Nikiforov_1988}, which is what we want to demonstrate in this note. Surprisingly, in the extensive literature on this topic, we have so far found only two papers \cite{Melnikov_1985,Ivashchuk_1996} where this method is mentioned in connection with similar problems, but in our opinion it is not used in the most optimal way.

\section{The Nikiforov-Uvarov method}
Second-order differential equations of generalized hypergeometric type, which have the following form 
\begin{equation}
u^{\prime\prime}+\frac{\pi_1(\mathrm{z})}{\sigma(\mathrm{z})}\,u^\prime+\frac{\sigma_1(\mathrm{z})}{\sigma^2(\mathrm{z})}\,u=0,
\label{eq4} 
\end{equation}
can be solved using the Nikiforov-Uvarov technique. In (\ref{eq4}), prime indicates differentiation with respect to the independent variable $\mathrm{z}$ (which may be complex); $\sigma(\mathrm{z})$ and $\sigma_1(\mathrm{z})$ are polynomials, at most of second degree, and $\pi_1(\mathrm{z})$ is a polynomial, at most of first degree.
Many problems in quantum mechanics can lead to equations of this type, and the Nikiforov-Uvarov method has been widely used in this context \cite{Ellis_2023,Tezcan_2009,Suslov_2020,Berkdemir_2012}. Recently, the Nikiforov-Uvarov method was used in the study of analytically solvable model of flyby-induced gravitational displacement memory
effect \cite{Zhang_2025}. Below, while describing the Nikiforov-Uvarov approach, we follow this last article, but for the benefit of the reader we offer a considerably more extensive description.

The first step is to realize that the set of solutions to equation (\ref{eq4}) is invariant under a certain kind of ``gauge" transformation
\begin{equation}
u(\mathrm{z})=e^{\varphi(\mathrm{z})}y(\mathrm{z}),    
\label{eq5}
\end{equation}
if 
\begin{equation}
\varphi^\prime=\frac{\pi(\mathrm{z})}{\sigma(\mathrm{z})},
\label{eq7}
\end{equation}
where $\pi(\mathrm{z})$ is some polynomial, at most of first degree. In this case, the function $y(\mathrm{z})$ is also of generalized hypergeometric type, since it will satisfy the equation
\begin{equation}
y^{\prime\prime}+\frac{\tau(\mathrm{z})}{\sigma(\mathrm{z})}y^\prime+\frac{\sigma_2(\mathrm{z})}{\sigma^2(\mathrm{z})}y=0,
\label{eq8}    
\end{equation}
where
\begin{equation}
\tau(\mathrm{z})=\pi_1(\mathrm{z})+2\pi(\mathrm{z})
\label{eq9}
\end{equation}
is a polynomial, at most of first degree, and
\begin{equation}
\sigma_2(\mathrm{z})=\sigma_1(\mathrm{z})+\pi^2(\mathrm{z})+\pi(\mathrm{z})\left [\pi_1(\mathrm{z})-\sigma^\prime(\mathrm{z})\right]+\pi^\prime(\mathrm{z})\sigma(\mathrm{z})
\label{eq10}
\end{equation}
is a polynomial, at most of second degree. 

The equation (\ref{eq4}) can be simplified by utilizing the freedom in selecting the polynomial $\pi(\mathrm{z})$. In particular, the equation (\ref{eq8}) reduces to an equation of hypergeometric type 
\begin{equation}
\sigma(\mathrm{z})y^{\prime\prime}+\tau(\mathrm{z})y^\prime+\lambda y=0,
\label{eq12}
\end{equation}
if $\pi(\mathrm{z})$ is selected so that 
\begin{equation} 
\sigma_2(\mathrm{z})=\lambda\sigma(\mathrm{z}), 
\label{eq11} 
\end{equation} 
where $\lambda$ is a constant.
Equations (\ref{eq10}) and (\ref{eq11}) imply that $\pi(\mathrm{z})$ is the root of a quadratic equation and has the following form
\begin{equation}
\pi(\mathrm{z})=\frac{\sigma^\prime-\pi_1}{2}\pm\sqrt{\left(\frac{\sigma^\prime-\pi_1}{2}\right)^2-\sigma_1+k\sigma},
\label{eq15}
\end{equation}
where $k = \lambda - \pi^\prime$. Only if 
\begin{equation} 
\sigma_3(\mathrm{z})=\left(\frac{\sigma^\prime-\pi_1}{2}\right)^2-\sigma_1+k\sigma \label{eq16} 
\end{equation} 
is the square of a first degree polynomial, will $\pi(\mathrm{z})$ be a polynomial. In this case, $\sigma_3(\mathrm{z})$ has a double root and its discriminant is zero:
\begin{equation}
\Delta(\sigma_3)=0.
\label{eq17}
\end{equation}
Equation (\ref{eq17}) determines the constant $k$ and hence the constant $\lambda$.

It is interesting that from the point of view of algebraic geometry, finding the polynomial $\pi(\mathrm{z})$ corresponds to the primary decomposition of the ideal \cite{Kikuchi_2020}. In general, there will be four possible solutions for the combination $(\pi,k)$ (there will be four primary ideals), and only one will be physically acceptable.

In bound state problems, normalization of the wave function requires polynomial solutions of the hypergeometric type equation (\ref{eq12}). Such solutions exist only for certain values of $\lambda$. The corresponding ``quantization condition" can be obtained in the following way. The derivatives of the function $y(\mathrm{z})$, $v_n(\mathrm{z})=y^{(n)}(\mathrm{z})$, $n=0,1,2,\ldots$, also satisfy hypergeometric type equation:
\begin{equation}
\sigma(\mathrm{z}) v_n^{\prime\prime}+\tau_n(\mathrm{z}) v_n^\prime +\mu_n(\mathrm{z}) v_n=0,
\label{eq20}
\end{equation}
where, as can be proved by induction,
\begin{equation}
\tau_n(\mathrm{z})=n\sigma^\prime+\tau,\;\;\;\mu_n(\mathrm{z})=\lambda+n\tau^\prime+\frac{1}{2}n(n-1)\sigma^{\prime\prime}.
\label{eq23}
\end{equation}
If $y(\mathrm{z})=y_n(\mathrm{z})$ is a polynomial of degree $n$, then $v_n$ is a constant and the equation (\ref{eq20}) will be fulfilled only if $\mu_n=0$. Therefore, we will have polynomial solutions only for 
\begin{equation}
\lambda=\lambda_n=- n\tau^\prime-\frac{1}{2}n(n-1)\sigma^{\prime\prime}.
\label{eq24}
\end{equation}
To find the corresponding polynomial $y_n(\mathrm{z})$, we first introduce the weight functions $\rho_m(\mathrm{z})$, $m=0,1,\ldots n$, that satisfy Pearson equation \cite{Ismail_2005} 
\begin{equation}
(\sigma\rho_m)^\prime=\rho_m\tau_m.
\label{eq25}
\end{equation}
Then $\rho_m\tau_m v_m^\prime+\sigma\rho_m v_m^{\prime\prime}=(\sigma\rho_m v_m^\prime)^\prime$ and equations (\ref{eq20}) can be rewritten in the self-adjoint form
\begin{equation}
(\sigma\rho_m v_{m+1})^\prime+\mu_m\rho_m v_m=0,
\label{eq26}
\end{equation}
since $v_m^\prime=v_{m+1}$. From equation (\ref{eq25}) we have ($\rho_0(\mathrm{z}) \equiv \rho(\mathrm{z})$)
\begin{equation}
\frac{(\sigma\rho_m)^\prime}{\rho_m}-\frac{(\sigma\rho)^\prime}{\rho}=\sigma\left (\frac{\rho_m^\prime}{\rho_m}-\frac{\rho}{\rho}\right )=\tau_m-\tau=m\sigma^\prime,
\label{eq27}
\end{equation}
where in the last step we used equation (\ref{eq23}). Dividing equation (\ref{eq27}) by $\sigma$ and integrating, we get (up to an unimportant constant factor)
\begin{equation}
\rho_m(\mathrm{z})=\sigma^m(\mathrm{z})\rho(\mathrm{z}),\;\;m=0,1,\ldots n.
\label{eq28}
\end{equation}
Therefore $\sigma\rho_m=\rho_{m+1}$, $m=0,1,\ldots n-1$, and equation (\ref{eq26}) can be rewritten as follows:
\begin{equation}
\rho_m v_m=-\frac{1}{\mu_m}(\rho_{m+1}v_{m+1})^\prime,\;\; m=0,1,\ldots n-1.
\label{eq29}
\end{equation}
By repeated use of this relation, we get
\begin{eqnarray}
&& \rho v_0=-\frac{1}{\mu_0}(\rho_1 v_1)^\prime \nonumber \\
&& =\left (-\frac{1}{\mu_0}\right )\left (-\frac{1}{\mu_1}\right )(\rho_2 v_2)^{\prime\prime}=\cdots=\frac{(-1)^n}{\mu_0\mu_1\cdots \mu_{n-1}}(\rho_n v_n)^{(n)},
\label{eq30}
\end{eqnarray} 
and since $v_0=y_n$, and $v_n$ is a constant, we get the Rodrigues formula for the polynomial $y_n(\mathrm{z})$:
\begin{equation}
y_n(\mathrm{z})=\frac{B_n}{\rho(\mathrm{z})}[\rho_n(\mathrm{z})]^{(n)}=\frac{B_n}{\rho(\mathrm{z})}\left[\sigma^n(\mathrm{z})\rho(\mathrm{z})\right]^{(n)},
\label{eq31}
\end{equation}
where $B_n$ is some (normalization) constant.

\section{Coulomb potential in spaces of constant curvature}
The Coulomb potential created by a charge density $\rho(r)$ is determined by the Poisson equation
\begin{equation}
    \Delta \phi(r)=-4\pi \rho(r),
\label{eq3-1}    
\end{equation}
where, in spaces of constant curvature, the Laplace-Beltrami operator has the form
\begin{equation}
    \Delta=\frac{1}{\sqrt{|g|}}\frac{\partial}{\partial x^i}\left (\sqrt{|g|}g^{ij}\frac{\partial}{\partial x^j}\right).
\label{eq3-2}    
\end{equation}
For unified description of constant curvature spaces, it is convenient to introduce generalized sine and cosine functions \cite{Ballesteros_1993}:
\begin{equation}
 S_\kappa(\mathrm{z})=\left \{\begin{array}{l} \frac{1}{\sqrt{\kappa}}\sin{(\sqrt{\kappa}\mathrm{z})},\;\mathrm{if}\;\kappa>0,\\
\mathrm{z},\;\mathrm{if}\;\kappa=0,\\\frac{1}{\sqrt{-\kappa}}\sinh{(\sqrt{-\kappa}\mathrm{z})},\;\mathrm{if}\;\kappa<0,\end{array}\right .\;\;\; C_\kappa(\mathrm{z})=\left \{\begin{array}{l} \cos{(\sqrt{\kappa}\mathrm{z})},\;\mathrm{if}\;\kappa>0,\\
1,\;\mathrm{if}\;\kappa=0,\\ \cosh{(\sqrt{-\kappa}\mathrm{z})},\;\mathrm{if}\;\kappa<0,\end{array}\right . 
\label{eq3-3}
\end{equation}
which satisfy
\begin{eqnarray}
&& S_\kappa(\mathrm{z})=\frac{e^{i\sqrt{\kappa}\mathrm{z}}-e^{-i\sqrt{\kappa}\mathrm{z}}}{2i\sqrt{\kappa}}, \; C_\kappa(\mathrm{z})=\frac{e^{i\sqrt{\kappa}\mathrm{z}}+e^{-i\sqrt{\kappa}\mathrm{z}}}{2}, \;
\nonumber \\
&& C_\kappa^2+\kappa S_\kappa^2=1, \;\;\;\;\; S_\kappa^\prime=C_\kappa,\;\;\;\;\;  C_\kappa^\prime=-\kappa S_\kappa.
\label{eq3-4}
\end{eqnarray} 
The generalized tangent is determined in a natural way as $T_\kappa(\mathrm{z})=\frac{S_\kappa(\mathrm{z})}{C_\kappa(\mathrm{z})}$, and satisfies
\begin{equation}
1+\kappa T_\kappa^2=\frac{1}{C_\kappa^2},\;\kappa+\frac{1}{T_\kappa^2}=\frac{1}{S_\kappa^2}.
\label{eq3-5}
\end{equation}
Using generalized trigonometric functions, line element in spaces of constant curvature can be expressed in the form \cite{Wald_1984}
\begin{equation}
    dl^2=dr^2+S^2_\kappa(r)\left(d\theta^2+\sin^2{\theta}\,d\varphi^2\right),
    \label{eq3-6}
\end{equation}
which corresponds to the metric tensor $g_{ij}=\mathrm{diag}(1,S_\kappa^2(r),S_\kappa^2(r)\sin^2{\theta})$. Then equation (\ref{eq3-1}) outside the point-source takes the form
\begin{equation}
    \frac{1}{S_\kappa^2(r)}\frac{d}{dr}\left (S_\kappa^2(r)\frac{d\phi}{dr}\right)=0,
    \label{eq3-7}
\end{equation}
which can be easily solved. The solution is determined up to a constant factor, and this latter is determined by the requirement that in the limit of flat space $\kappa\to 0$ we obtain the usual Coulomb potential. Finally, the Coulomb potential energy for hydrogen-like elementary atoms in spaces of constant curvature turns out to be equal to
\begin{equation}
    V(r)=-\frac{e^2}{T_\kappa(r)}.
    \label{eq3-8}
\end{equation}
The analog of Newton's force law for the hyperbolic space $\mathbb{H}^3$, corresponding to (\ref{eq3-8}), was proposed already by Lobachevski \cite{Shchepetilov_2005}.
 
\section{Elementary atoms in spaces of constant curvature}
The Schr\"{o}dinger equation for elementary atoms in spaces of constant curvature has the form 
\begin{equation}
\left[-\Delta+\frac{2m}{\hbar^2}V\right]\Psi=\frac{2m}{\hbar^2}E\Psi,
\label{eq4-1}
\end{equation}
where $\Delta$ is the Laplace-Beltrami operator given by (\ref{eq3-2}) and for elementary atoms the potential energy $V(r)$ has the form (\ref{eq3-8}). Because of rotational symmetry inherent in (\ref{eq4-1}) for the central potential (\ref{eq3-8}), the angular part of the wave function $\Psi$ is the ordinary spherical function: $\Psi=G(r)Y_{lm}(\theta,\varphi)$. As for the radial part $G(r)$, it satisfies the differential equation \cite{Nieto_1999}
\begin{equation}
\left [-\frac{1}{S_\kappa^2(r)}\frac{d}{dr}\left (S_\kappa^2(r)\frac{d}{dr}\right)+\frac{l(l+1)}{S_\kappa^2(r)}+\frac{2m}{\hbar^2}V-\frac{2m}{\hbar^2}E\right ]G(r)=0.    
\label{eq4-2}
\end{equation}
It is convenient to introduce a dimensionless variable
\begin{equation}
  z=\frac{1}{\sqrt{\kappa}\,T_\kappa(r)}.  
\label{4-3}
\end{equation}
Then the radial equation takes the form
\begin{equation}
\frac{d^2G}{dz^2}+\frac{\lambda_E+\beta_Rz-l(l+1)(1+z^2)}{(1+z^2)^2}\,G=0, 
\label{eq4-3}
\end{equation}
where
\begin{equation}
\lambda_E=\frac{2mE}{\hbar^2\kappa},\;\beta_R=\frac{2me^2}{\hbar^2\sqrt{\kappa}}.   
\label{eq4-4} 
\end{equation}
Equation (\ref{eq4-3}) is of generalized hypergeometric type with
\begin{equation}
\sigma=1+z^2,\;\;\pi_1=0,\;\;\sigma_1=\lambda_E+\beta_R z-l(l+1)(1+z^2). 
\label{eq4-5}
\end{equation}
Therefore, according to (\ref{eq16}),
\begin{equation}
\sigma_3=z^2[1+k+l(l+1)]-\beta_R z+k+l(l+1)-\lambda_E,
\label{eq4-6}
\end{equation}
and $\Delta(\sigma_3)=0$ condition will give
\begin{equation}
\beta_R^2-4[1+k+l(l+1)][k+l(l+1)-\lambda_E]=0.
\label{eq4-7}
\end{equation}
For convenience, let us introduce $x=1+k+l(l+1)$. Then
\begin{equation}
\frac{\beta_R^2}{4}=x(x-1-\lambda_E). 
\label{eq4-8}
\end{equation}
Of the two possible $\pi$, we choose the one with $\pi^\prime<0$ to obtain a physically acceptable wave function that decreases with increasing $r$ \cite{Berkdemir_2012}. This choice leads to
\begin{equation}
\pi=(1-\sqrt{x})z+\sqrt{x-1-\lambda_E}. 
\label{eq4-9}
\end{equation}
Then, since $\tau=2\pi$, the quantization condition (\ref{eq24}) takes the form
\begin{equation}
    \lambda=- n_r\tau^\prime-\frac{1}{2}n_r(n_r-1)\sigma^{\prime\prime}=-2n_r(1-\sqrt{x})-n_r(n_r-1),
    \label{eq4-10}
\end{equation}
where $n_r=0,1,\ldots$ is the radial quantum number. If $k$ is expressed through $x$ in $k=\lambda-\pi^\prime=\lambda-1+\sqrt{x}$, then by recalling from the definition of $x$ that $k=x-1-l(l+1)$, we get a quadratic equation for $\sqrt{x}$:
\begin{equation}
    x-(2n_r+1)\sqrt{x}+n_r(n_r+1)-l(l+1)=0.
    \label{eq4-11}
\end{equation}
The solution that corresponds to the correct flat limit (not restricted by the  condition $n_r\ge l$) is $\sqrt{x}=n_r+l+1$. Therefore, from (\ref{eq4-8}), $\frac{\beta_R^2}{4}=n^2(n^2-1-\lambda_E)$ and 
\begin{equation}
 \lambda_E=n^2-1-\frac{\beta_R^2}{4n^2},
 \label{eq4-12}
\end{equation}
where $n=n_r+l+1$ is the principal quantum number. Using (\ref{eq4-4}), this implies the energy spectrum
\begin{equation}
    E_n=Ry\left(-\frac{1}{n^2}+(n^2-1)\,\kappa a_B^2\right ),\;\;\;a_B=\frac{\hbar^2}{me^2},\;\;\;Ry=\frac{\hbar^2}{2ma_B^2}
=\frac{me^4}{2\hbar^2},
\label{eq4-13}
\end{equation}
first found by Schr\"{o}dinger \cite{Schrodinger_1940} in case of $\mathbb{S}^3$, and by Infeld and Schild \cite{Infeld_1945} in case of $\mathbb{H}^3$.

For the gauge function $\varphi$ we have the equation
\begin{equation}
 \frac{d\varphi}{dz}=\frac{\pi}{\sigma}=\frac{-(n-1)z+\sqrt{n^2-1-\lambda_E}}{1+z^2}=\frac{-(n-1)z+\frac{\beta_R}{2n}}{1+z^2},
 \label{eq4-14}
\end{equation}
which can be easily solved with the result
\begin{equation}
    \varphi(z)=-\frac{1}{2}(n-1)\ln{(1+z^2)}+\frac{\beta_R}{2n}\arctan{z}.
    \label{eq4-15}
\end{equation}

Using $1+z^2=\frac{1}{\kappa S^2_\kappa(r)}$ and $\arctan{z}=\frac{i}{2}\ln{\frac{1-iz}{1+iz}}$ in the first and second terms respectively, the gauge function $\varphi(z)$ can be rewritten as:
\begin{equation}
\varphi(z)=
\ln{(\sqrt{\kappa}S_\kappa(r))^{n-1}}+\frac{\pi}{2n\,\sqrt{\kappa}\, a_B}-\frac{1}{n}\,\frac{r}{a_B},
 \label{eq4-16}
\end{equation}
where the second term contributes only to the normalization of the wave function. Therefore, ultimately, based on (\ref{eq5}), the radial wave function takes the form
\begin{equation}
 G_{n,n_r}(r)=B_{n,n_r}[\sqrt{\kappa}S_\kappa(r)]^{n-1}e^{-\frac{1}{n}\,\frac{r}{a_B}}y_{n_r}^n(z),
 \label{eq4-17}
\end{equation}
where $B_{n,n_r}$ is a normalization constant, $n_r=n-l-1$, and the polynomials $y_{n_r}^n(z)$ are given by the Rodrigues formula (\ref{eq31}), with the assumption that $(-1)^{n_r}[\mu_0\mu_1\cdots \mu_{nr-1}]^{-1}v_{n_r}=1$ (i.e. without the normalization constant, since it is already present in (\ref{eq4-17})). For them we need the weight functions $\rho(z)$ determined from the equation $\frac{d}{dz}(\rho\sigma)=\rho\tau$, which implies $\frac{(\rho\sigma)^\prime}{\rho\sigma}=\frac{\tau}{\sigma}=2\varphi^\prime$. Hence
\begin{equation}
\rho(z)=\frac{1}{1+z^2}\,e^{2\varphi}=(1+z^2)^{-n}e^{\frac{\beta_R}{n}\,\arctan{z}},
\label{eq4-18}
\end{equation}
and
\begin{equation}
y_{n_r}^n(z)=\frac{1}{\rho(z)}\left[\,(1+z^2)^{n_r}\,\rho(z)\,\right]^{(n_r)}.
\label{eq4-19}
\end{equation}
Comparing to the definition of Romanovsky polynomials $R_n^{(\alpha,\beta)}(z)$ through the weight function $\rho(z)=(1+z^2)^{\beta-1}e^{-\alpha\,\arccot{z}}$ \cite{Raposo_2007,Kirchbach_2008}, we see that up to normalization $y_{n_r}^n(z)=R_{\;n_r}^{(\frac{\beta_R}{n},\,1-n )}(z)$. Expressing $\arctan(z)$ again through the logarithm and using  $\beta_{R} = \frac{2}{a_{B}\sqrt{\kappa}}$, the weight function can be reduced to the form $\rho(z)=(1+iz)^{-n-\frac{i}{n}\frac{1}{a_B\sqrt{\kappa}}}\;(1-iz)^{-n+\frac{i}{n}\frac{1}{a_B\sqrt{\kappa}}}$. Comparing with the definition of Jacobi polynomials
\begin{equation}
P_n^{(\alpha,\beta)}(z)=\frac{(-1)^n}{2^nn!}(1-z)^{-\alpha}(1+z)^{-\beta} \frac{d^n}{dz^n}\left\{(1-z)^\alpha (1+z)^\beta (1-z^2)^n\right \},
\label{eq4-20} 
\end{equation}
we see that $y_{n_r}^n(z)$ can be expressed in terms of complex Jacobi polynomials \cite{Bessis_1979}
\begin{equation}
y_{n_r}^n(z)=(-i)^{n_r}\,2^{n_r}\,n_r!\,P_{n_r}^{(\alpha,\beta)}(iz),\;\;\alpha=-n+\frac{i}{n}\frac{1}{a_B\sqrt{\kappa}},\;\;\beta=-n-\frac{i}{n}\frac{1}{a_B\sqrt{\kappa}}. 
\label{eq4-21}
\end{equation}
Using (\ref{eq4-21}) and three-term recurrence relation for Jacobi polynomials (see, for example, \cite{Ismail_2005})
\begin{eqnarray} &
P_{\;n+1}^{(\alpha,\,\beta)}(z)=
& \nonumber \\ &
\left (\frac{(2n+\alpha+\beta+1)(2n+\alpha+\beta+2)}
{2(n+1)(n+\alpha+\beta+1)}\,z+\frac{(\alpha^2-\beta^2)(2n+\alpha+\beta+1)}
{2(n+1)(n+\alpha+\beta+1)(2n+\alpha+\beta)}\right )P_{\;n}^{(\alpha,\,\beta)}(z)\nonumber \\ & -
\frac{(n+\alpha)(n+\beta)(2n+\alpha+\beta+2)}{(n+1)(n+\alpha+\beta+1)
(2n+\alpha+\beta)}\,P_{\;n-1}^{(\alpha,\,\beta)}(z), &
\label{eq4-22}
\end{eqnarray}
we can obtain the corresponding three-term recurrence relation for the polynomials $y_{n_r}^n(z)$, convenient for their numerical calculation:
\begin{eqnarray} &
y_{n_r}^n(z)=\frac{2n+1-2n_r}{2n-n_r}\left[\frac{\beta_R}{n+1-n_r}-2(n-n_r)z\right]y_{n_r-1}^n(z) & \nonumber \\ & -
\frac{(n_r-1)(n-n_r)}{2n-n_r}\left[4(n+1-n_r)+\frac{\beta_R^2}{n^2(n+1-n_r)}\right]y_{n_r-2}^n(z),&
\label{eq4-23}
\end{eqnarray}
with 
\begin{equation}
y_0^n(z)=1,\;\;y_1^n(z)=2(1-n)z+\frac{\beta_R}{n}    
\label{eq4-24}
\end{equation}
as the initial values for the recurrence.

Another recurrence relation for the polynomials $y_{n_r}^n(z)$, containing the derivative operator, can be obtained as follows \cite{Raposo_2007}. It is clear from (\ref{eq29}) that
\begin{eqnarray}
&&\rho_1 v_1=-\frac{1}{\mu_1}(\rho_2 v_2)^\prime
\nonumber \\ &&=\left (-\frac{1}{\mu_1}\right )\left (-\frac{1}{\mu_2}\right )(\rho_3 v_3)^{\prime\prime}=\cdots=\frac{(-1)^{n-1}}{\mu_1\mu_2\cdots \mu_{n-1}}(\rho_n v_n)^{(n-1)}.
\label{eq4-25}
\end{eqnarray}
Comparing with (\ref{eq30}) and (\ref{eq31}), and remembering that $$\frac{(-1)^{n_r}}{\mu_0\mu_1\cdots \mu_{nr-1}}\;v_{n_r}=1,$$ we get
\begin{eqnarray}
&& \frac{dy_{n_r}^n(z)}{dz}=-\mu_0\frac{1}{\rho(z)\sigma(z)}\left[\sigma^{n_r}(z)\rho(z)\right]^{(n_r-1)} \nonumber \\
&& =-n_r(2n-n_r-1)\frac{1}{\rho(z)\sigma(z)}\left[\sigma^{n_r}(z)\rho(z)\right]^{(n_r-1)},
\label{eq4-26}
\end{eqnarray}
where in the last step we have used $\mu_0=\lambda_{n_r}=n_r(2n-n_r-1)$ which follows from (\ref{eq4-10}).

Since $(\rho\sigma)^\prime=\rho\tau$, $(\rho\sigma\sigma^{n_r})^\prime=(\tau+n_r\sigma^\prime)\rho\sigma^{n_r}$ and 
\begin{eqnarray}
&&
y_{n_r+1}^n=\frac{1}{\rho}\left[\rho\sigma^{n_r+1}\right ]^{(n_r+1)}=\frac{1}{\rho}\frac{d^{n_r}}{dz^{n_r}}\left[(\tau+n_r\sigma^\prime)(\rho\sigma^{n_r})\right ] \nonumber \\
&& =\frac{\tau+n_r\sigma^\prime}{\rho}\left[\rho\sigma^{n_r}\right]^{(n_r)}+ \frac{n_r(\tau^\prime+n_r\sigma^{\prime\prime})}{\rho}\left[\rho\sigma^{n_r}\right]^{(n_r-1)},
\label{eq4-27}
\end{eqnarray}
where in the last step we used Leibniz's formula for multiple derivatives of a product. In the first term we recognize $y_{n_r}^n$, while the second term can be expressed in terms of $(y_{n_r}^n)^\prime$ according to (\ref{eq4-26}). As a result we get
$$y_{n_r+1}^n(z)=(\tau+n_r\sigma^\prime)\,y_{n_r}^n(z)-\frac{\sigma(\tau^\prime+n_r\sigma^{\prime\prime})}{2n-n_r-1}\frac{dy_{n_r}^n(z)}{dz},$$
or
\begin{equation}
y_{n_r+1}^n(z)=\left[2(1-n+n_r)z+\frac{\beta_R}{n}\right] y_{n_r}^n(z)-\frac{2(1+z^2)(1-n+n_r)}{2n-n_r-1} \frac{dy_{n_r}^n(z)}{dz}.
\label{eq4-28}
\end{equation}

\section{Normalization and flat limit}
To determine the normalization factor $B_{n,n_r}$ in (\ref{eq4-17}) one can calculate the corresponding normalization integrals directly \cite{Vinitsky_1993}. Since this is a rather laborious approach, we prefer the indirect way using raising and lowering operators \cite{Leemon_1978,Higgs_1979}. As a first step, we construct raising $\hat a_+$ and lowering $\hat a_-$ operators for the polynomials $y_{n_r}^n(z)$ such that
\begin{equation}
 \hat a_+\,y_{n_r}^n(z)= y_{n_r+1}^n(z),\;\;\;  \hat a_-\,y_{n_r}^n(z)= y_{n_r-1}^n(z).
 \label{eq5-1}
\end{equation}
The raising operator $\hat a_+$ can be extracted  from (\ref{eq4-28}):
\begin{equation}
\hat a_+=(n-n_r-1)\left [-2z+\frac{\beta_R}{n(n-n_r-1)}+\frac{2(1+z^2)}{2n-n_r-1}\frac{d}{dz}\right],
\label{eq5-2}
\end{equation}
while the lowering operator $\hat a_-$ is obtained if (\ref{eq4-28}) is combined with the recurrence relation (\ref{eq4-23}). The result is
\begin{equation}
\hat a_-=\frac{n^2(n-n_r)}{4n^2(n-n_r)^2+\beta_R^2}\left [2z+\frac{\beta_R}{n(n-n_r)}-\frac{2(1+z^2)}{n_r}\frac{d}{dz}\right].
\label{eq5-3}
\end{equation}
Using $n_r=n-l-1$ and
\begin{equation}
\frac{d}{dr}=\frac{dz}{dr}\,\frac{d}{dz}=-\frac{1}{\sqrt{\kappa}}\,\frac{1}{S_\kappa^2}\,\frac{d}{dz}=-\sqrt{\kappa}\,(1+z^2)\,\frac{d}{dz},
\label{eq5-4}
\end{equation}
we can rewrite (\ref{eq5-2}) and (\ref{eq5-3}) in terms of the radial variable $r$ as follows:
\begin{eqnarray}
&& \hat a_+=\frac{2}{\sqrt{\kappa}}\left [-\frac{l}{T_\kappa(r)}+\frac{1}{na_B}-\frac{l}{n+l}\,\frac{d}{dr}\right],\;\; \nonumber \\
&& \hat a_-=\frac{\sqrt{\kappa}}{2}\,\frac{n^2a_B^2}{1+n^2(l+1)^2a_B^2\kappa}\,\left [\frac{l+1}{T_\kappa(r)}+\frac{1}{na_B}+\frac{l+1}{n-l-1}\,\frac{d}{dr}\right].
\label{eq5-5}
\end{eqnarray}
To raise $\hat{a}_+$ and $\hat{a}_-$ to raising $\hat{A}_+$ and lowering $\hat{A}_-$ operators for (unnormalized) radial wave functions:
\begin{eqnarray}
&& \hat{A}_+\,[\sqrt{\kappa} S_\kappa]^{n-1}\,e^{-\frac{r}{na_B}}\,y_{n_r}^n=[\sqrt{\kappa} S_\kappa]^{n-1}\,e^{-\frac{r}{na_B}}\,y_{n_r+1}^n,\;\;\; \nonumber \\
&& \hat{A}_-\,[\sqrt{\kappa} S_\kappa]^{n-1}\,e^{-\frac{r}{na_B}}\,y_{n_r}^n=[\sqrt{\kappa} S_\kappa]^{n-1}\,e^{-\frac{r}{na_B}}\,y_{n_r-1}^n,\;\;\;
\label{eq5-6}
\end{eqnarray}
we take $\hat{A}_+=\hat{a}_++f_+(r)$, $\hat{A}_-=\hat{a}_-+f_-(r)$, where the role of the functions $f_+$ and $f_-$ is to compensate for the action of the derivative operator $\frac{d}{dr}$ on $[\sqrt{\kappa} S_\kappa]^{n-1}\,e^{-\frac{r}{na_B}}$. A simple calculation will give
\begin{eqnarray}
&& f_+(r)=\frac{2}{\sqrt{\kappa}}\,\frac{l}{n+l}\,\left[\frac{n-1}{T_\kappa(r)}-\frac{1}{na_B}\right],\;\;\; \nonumber \\
&& f_-(r)=-\frac{\sqrt{\kappa}}{2}\,\frac{n^2a_B^2}{1+n^2(l+1)^2a_B^2\kappa} \,\frac{l+1}{n-l-1}\,\left[\frac{n-1}{T_\kappa(r)}-\frac{1}{na_B}\right],
\label{eq5-7}
\end{eqnarray}
and consequently we get
\begin{eqnarray} 
&& \hspace*{-10mm}\hat{A}_+(l)= \frac{2}{\sqrt{\kappa}}\, \frac{l}{n+l}\, \left[-\frac{l+1}{T_\kappa(r)}+\frac{1}{la_B}-\frac{d}{dr}\right],\;\; \nonumber \\
&& \hspace*{-10mm}\hat{A}_-(l)=\frac{\sqrt{\kappa}}{2}\,  
\frac{n^2a_B^2}{1+n^2(l+1)^2a_B^2\kappa}\, \frac{l+1}{n-l-1}\, \left[-\frac{l}{T_\kappa(r)}+\frac{1}{(l+1)a_B}+\frac{d}{dr}\right].
\label{eq5-8}
\end{eqnarray}
The operators in parentheses are actually the lowering and raising operators (with respect to the quantum number $l$) used in \cite{Leemon_1978}. Writing $\hat{A}_+(l)=\alpha_l\hat{d}_+(l)$ and $\hat{A}_-(l)=\beta_l\hat{d}_-(l)$ with
\begin{equation}
\alpha_l=\frac{2}{\sqrt{\kappa}}\,\frac{l}{n+l},\;\;\;
\beta_l=\frac{\sqrt{\kappa}}{2}\,\frac{n^2a_B^2}{1+n^2(l+1)^2a_B^2\kappa} \,\frac{l+1}{n-l-1},
\label{eq5-9}
\end{equation}
we find
\begin{equation}
\hat d_+(l+1)\hat d_-(l)=-\frac{d^2}{dr^2}-\frac{2}{T_\kappa}\frac{d}{dr}+\frac{l(l+1)}{S_\kappa^2}-\frac{2}{a_BT_\kappa}+\frac{1}{(l+1)^2a_B^2}-l(l+2)\kappa.    
\label{eq5-10}
\end{equation}
In derivation of (\ref{eq5-10}) we have used
$$\frac{d}{dr}\frac{1}{T_\kappa}=\frac{1}{T_\kappa}\frac{d}{dr}-\frac{1}{S_\kappa^2}\;\;\mathrm{and}\;\;\frac{1}{T_\kappa^2}=\frac{1}{S_\kappa^2}-\kappa.$$
Since the (unnormalized) quantum state $|n,n_r\rangle=[\sqrt{\kappa}S_\kappa(r)]^{n-1}e^{-\frac{1}{n}\,\frac{r}{a_B}}y_{n_r}^n(z)$ satisfies the Schr\"{o}dinger equation (\ref{eq4-2}), we have
\begin{eqnarray}
&& \hspace*{-5mm} \langle n,n_r|\hat d_+(l+1)\hat d_-(l)|n,n_r\rangle=\left[ \kappa[\lambda_E-l(l+2)]+\frac{1}{(l+1)^2a_B^2}\right] \,  \langle n,n_r|n,n_r\rangle \nonumber \\ 
&& \hspace*{-5mm} =\frac{[n^2-(l+1)^2][1+n^2(l+1)^2\kappa a_B^2]}{n^2(l+1)^2a_B^2} \, \langle n,n_r|n,n_r\rangle \ge 0.
\label{eq5-11}    
\end{eqnarray}
Therefore, $\hat d_+(l+1)\hat d_-(l)$ is a positive and hence self-adjoint operator, which means that $\hat d^+_-(l)=\hat d_+(l+1)$ \cite{Leemon_1978}. Since $|n,n_r-1\rangle=\hat A_-(l)|n,n_r\rangle=\beta_l \hat d_-(l)|n,n_r\rangle$ and $\langle n,n_r|\hat A_+(l+1)\hat A_-(l)|n,n_r\rangle=\langle n,n_r|n,n_r\rangle$, we get
\begin{eqnarray}
&& \langle n,n_r-1|n,n_r-1\rangle=\beta_l^2\langle n,n_r|\hat d_+(l+1)\hat d_-(l)|n,n_r\rangle \nonumber \\
&& =\frac{\beta_l}{\alpha_{l+1}}\langle n,n_r|\hat A_+(l+1)\hat A_-(l)|n,n_r\rangle=\frac{\beta_l}{\alpha_{l+1}}\langle n,n_r|n,n_r\rangle.
\label{eq5-12}
\end{eqnarray}
On the other hand, $B_{n,n_r}^{-2}=\langle n,n_r|n,n_r\rangle$. Therefore, it follows from (\ref{eq5-12}) that
\begin{equation}
B_{n,n_r}=\sqrt{\frac{\beta_l}{\alpha_{l+1}}}\,B_{n,n_r-1}=\frac{\sqrt{\kappa}}{2}\sqrt{\frac{n^2a_B^2}{1+n^2(l+1)^2\kappa a_B^2}\,\frac{n+l+1}{n-l-1}}\,B_{n,n_r-1}.
\label{eq5-13}
\end{equation}
This relation allows us to calculate the normalization coefficients recursively, starting from $B_{n,0}$. For the latter, we have
\begin{equation}
B_{n,0}^{-2}=\int S_\kappa^2(r)[\sqrt{\kappa}S_\kappa(r)]^{2(n-1)}e^{-\frac{2r}{na_B}}\,dr.
\label{eq5-14}
\end{equation}
If $\kappa<0$, then the integral in (\ref{eq5-14}) converges only for $n\sqrt{-\kappa}<\frac{1}{na_B}$, or $n^2<\frac{R}{a_B}$, where $R=\frac{1}{\sqrt{-k}}$. Therefore, in a space of constant negative curvature, hydrogen-like elementary atoms have only a finite, albeit very large $n\sim\sqrt{\frac{R}{a_B}}$, number of bound states \cite{Infeld_1945}.

In the flat limit $\kappa\to 0$, $S_\kappa(r)\to r$, 
$$B_{n,0}^{-2}\;\to\;\kappa^{n-1}\int\limits_0^\infty r^{2n}e^{-\frac{2r}{na_B}}\,dr=\kappa^{n-1}\left(\frac{na_B}{2}\right)^{2n+1}(2n)!,$$
and, as a result,
\begin{equation}
B_{n,n_r}\;\to\;(\sqrt{\kappa})^{n_r+1-n}\,\left(\frac{2}{na_B}\right)^{n+\frac{1}{2}-n_r}\,\sqrt{\frac{1}{2n(n-l-1)!(n+l)!}}.
\label{eq5-15}   
\end{equation}
On the other hand, when $\kappa\to 0$, then $$z\to\frac{1}{\sqrt{\kappa}\, r},\;\; 1+z^2\to\frac{1}{\kappa\, r^2}, \;\;\rho\to\kappa^n\,r^{2n}\,e^{\frac{\pi}{n\,\sqrt{\kappa}\, a_B}}\,e^{-\frac{2r}{na_B}},$$ and
\begin{eqnarray}
&& y_{n_r}^n\to \kappa^{-n_r}\,r^{-2n}\,e^{\frac{2r}{na_B}}\,\frac{d^{n_r}}{dz^{n_r}}\left[r^{2(n-n_r)}\,e^{-\frac{2r}{na_B}}\right] \nonumber \\ 
&& =(-1)^{n_r}\,(\sqrt{\kappa})^{-n_r}\,r^{-2n}\,\left(\frac{na_B}{2}\right)^{2n-n_r}e^{-1/t}\,\frac{d^{n_r}}{dt^{n_r}}\left [t^{-2(n-n_r)}e^{1/t}\right ],    
\label{eq5-16}
\end{eqnarray}
where $t=-\frac{na_B}{2r}$. Next we use the following equality \cite{Hadinger_1974,Duff_1949}
\begin{equation}
\frac{d^n}{dt^n}\left [t^{-k}\,e^{1/t}\right ]=(-1)^n\,n!\,t^{-(n+k)}\,e^{1/t}\,L_n^{k-1}\left(-\frac{1}{t}\right),
\label{eq5-17}
\end{equation}
where $L_n^m(x)$ is the associated Laguerre polynomial. As a result, (\ref{eq5-16}) takes the form
\begin{equation}
y_{n_r}^n\to (-1)^{n_r}\,(n-l-1)!\,(\sqrt{\kappa})^{-n_r}\,r^{-n_r}\,L_{n-l-1}^{2l+1}\left(\frac{2r}{na_B}\right ),
\label{eq5-18}
\end{equation}
and
\begin{equation}
|n,n_r\rangle \to (-1)^{n_r}\,(n-l-1)!\,(\sqrt{\kappa})^{n-1-n_r}\,r^{n-1-n_r}\,e^{-\frac{r}{na_B}}\,L_{n-l-1}^{2l+1}\left(\frac{2r}{na_B}\right ).
\label{eq5-19}
\end{equation}
Finally, combining (\ref{eq5-15}) and (\ref{eq5-19}), we get
\begin{equation}
B_{n,n_r}|n,n_r\rangle \to (-1)^{n-l-1}\,\frac{2}{n^2}\,\sqrt{\frac{(n-l-1)!}{a_B^3(n+l)!}}\,\left(\frac{2r}{na_B}\right)^l\,e^{-\frac{r}{na_B}}\,L_{n-l-1}^{2l+1}\left(\frac{2r}{na_B}\right ).
\label{eq5-20}
\end{equation} 
Up to a possible irrelevant sign, the right-hand side of (\ref{eq5-20}) is exactly the wave function of the hydrogen atom in flat space \cite{Arfken_2005}. Note that many different conventions are used for ordinary and associated Laguerre polynomials in the physics literature. We follow conventions of Arfken and Weber \cite{Arfken_2005}.

\section{Concluding remark}
The Nikiforov-Uvarov method has been used in many quantum mechanical problems \cite{Ellis_2023,Tezcan_2009,Suslov_2020,Berkdemir_2012}. Hydrogen-like atoms in spaces of constant curvature represent another quantum mechanical problem where this method can be successfully applied. Moreover, in our opinion, in this case the Nikiforov-Uvarov method provides the most natural and simple way to solve the problem.

\appendix
\section{Proof of the Duff identity}
We can prove the Duff identity (\ref{eq5-17}) by induction. If $n=0$, then the identity is true, since for all $m$, $L_0^m(x)=1$. Suppose it is true for some $n$. Then
\begin{eqnarray}
&& \frac{d^{n+1}}{dt^{n+1}}\left [t^{-k}e^{1/t} \right ]=(-1)^n\,n!\,\frac{d}{dt}\left[ t^{-(n+k)}\,e^{1/t}\,L_n^{k-1}\left(-\frac{1}{t}\right)\right] \nonumber \\ 
&& =(-1)^k\,n!\,x^2\,\frac{d}{dx}\left[ x^{n+k}\,e^{-x}\,L_n^{k-1}\left(x\right)\right],
\label{eqA-1}
\end{eqnarray}
where $x=-1/t$. Using the recurrence relation \cite{Arfken_2005}
\begin{equation}
x\frac{d}{dx}L_n^k(x)=n L_n^k(x)-(n+k) L_{n-1}^k(x),
\label{eqA-2}
\end{equation}
we get
\begin{eqnarray}
&& x^2\,\frac{d}{dx}\left[ x^{n+k}\,e^{-x}\,L_n^{k-1}\left(x\right)\right]= \nonumber \\
&& =x^{n+k+1}e^{-x}\left[(n+k-x)L_n^{k-1}(x)+x\frac{d}{dx}L_n^{k-1}(x)\right] \nonumber \\
&& =x^{n+k+1}e^{-x}\left[(2n+k-x)L_n^{k-1}(x)-(n+k-1)L_{n-1}^{k-1}(x)\right] \nonumber \\
&& =(n+1)\,x^{n+k+1}e^{-x}L_{n+1}^{k-1}(x),
\label{eqA-3}
\end{eqnarray}
where at the last step we have used another recurrence relation \cite{Arfken_2005}
\begin{equation}
(n+1)L_{n+1}^k(x)=(2n+k+1-x)L_n^k(x)-(n+k)L_{n-1}^k(x).
\label{eqA-4}    
\end{equation}
Returning to the original variable $t$ and substituting the result of (\ref{eqA-3}) into (\ref{eqA-1}), we get
\begin{equation}
\frac{d^{n+1}}{dt^{n+1}}\left [t^{-k}e^{1/t} \right ]=(-1)^{n+1}\,(n+1)!\,t^{-(n+k+1)}\,e^{1/t}\,L_{n+1}^{k-1}\left(-\frac{1}{t}\right).
\label{eqA-5}
\end{equation}
This ends the inductive step and thus the proof.

\bibliographystyle{elsarticle-num}
\bibliography{HA_Curved.bib}

\begin{thebibliography}{10}
\expandafter\ifx\csname url\endcsname\relax
  \def\url#1{\texttt{#1}}\fi
\expandafter\ifx\csname urlprefix\endcsname\relax\def\urlprefix{URL }\fi
\expandafter\ifx\csname href\endcsname\relax
  \def\href#1#2{#2} \def\path#1{#1}\fi

\bibitem{Gindikin_1988}
S.~G. Gindikin, Tales of Physicists and Mathematicians, Birkh{\"{a}}user, Boston, 1988.

\bibitem{Prasolov_2015}
V.~V. Prasolov, A.~B. Skopenkov, Some reflections on why {Lobachevsky} geometry was recognized, Mat. Prosveschenie 19~(Ser. 3) (2015) 99--108, in Russian.
\newblock \href {http://arxiv.org/abs/1307.4902} {\path{arXiv:1307.4902}}.

\bibitem{Boltianski_2002}
V.~Boltianski, A.~Savin, Conversations about Mathematics. Book 1: Discrete objects, Fima, MCNMO, Moscow, 2002, in Russian.

\bibitem{Gray_2007}
J.~Gray, Beltrami, {Klein}, and the acceptance of non-{Euclidean} geometry, in: Worlds Out of Nothing: A Course in the History of Geometry in the 19th Century, Springer London, London, 2007, pp. 219--232.
\newblock \href {https://doi.org/10.1007/978-1-84628-633-9_20} {\path{doi:10.1007/978-1-84628-633-9_20}}.

\bibitem{Chern_1979}
S.-S. Chern, From triangles to manifolds, Am. Math. Monthly 86~(5) (1979) 339--349.
\newblock \href {https://doi.org/10.1080/00029890.1979.11994807} {\path{doi:10.1080/00029890.1979.11994807}}.

\bibitem{Dombrovski_1991}
P.~Dombrovski, J.~Zitterbarth, {On the planetary motion in the 3-dim standard spaces $M^3_k$ of constant curvature $k\in \mathbb{R}$}, Demonstratio Mathematica 24~(3-4) (1991) 375--458.
\newblock \href {https://doi.org/doi:10.1515/dema-1991-3-405} {\path{doi:doi:10.1515/dema-1991-3-405}}.

\bibitem{Shchepetilov_2005}
A.~V. Shchepetilov, Comment on “central potentials on spaces of constant curvature: The {Kepler} problem on the two-dimensional sphere $\mathbb{S}^2$ and the hyperbolic plane $\mathbb{H}^2$”, J. Math. Phys. 46~(11) (2005) 114101.
\newblock \href {https://doi.org/10.1063/1.2107267} {\path{doi:10.1063/1.2107267}}.

\bibitem{Schrodinger_1940}
E.~Schr\"{o}dinger, \href{http://www.jstor.org/stable/20490744}{A method of determining quantum-mechanical eigenvalues and eigenfunctions}, Proc. Roy. Irish Acad. A 46 (1940) 9--16.
\newline\urlprefix\url{http://www.jstor.org/stable/20490744}

\bibitem{Shchepetilov_2006}
A.~V. Shchepetilov, Calculus and Mechanics on Two-Point Homogenous Riemannian Spaces, Springer-Verlag, Berlin, 2006.

\bibitem{Ovsiyuk_2025}
A.~V. Chichurin, E.~M. Ovsiyuk, V.~M. Red’kov, Problems in Quantum Mechanics and Field Theory with Mathematical Modelling, CRC Press, Boca Raton, 2025.

\bibitem{Borisov_2004}
A.~Borisov, I.~Mamaev (Eds.), Classical dynamics in non-Euclidean spaces, RCD, Moscow, 2004.

\bibitem{Nieto_1999}
L.~M. Nieto, M.~Santander, H.~C. Rosu, Hydrogen atom as an eigenvalue problem in 3-d spaces of constant curvature and minimal length, Mod. Phys. Lett. A 14~(35) (1999) 2463--2469.
\newblock \href {https://doi.org/10.1142/S021773239900256X} {\path{doi:10.1142/S021773239900256X}}.

\bibitem{Redkov:2011}
V.~M. Red'kov, E.~M. Ovsiyuk, {Parabolic Coordinates and the Hydrogen Atom in Spaces $H_{3}$ and $S_{3}$}, Nonlin. Phenom. Complex Syst. 14 (2011) 106--125.
\newblock \href {http://arxiv.org/abs/1108.6176} {\path{arXiv:1108.6176}}.

\bibitem{Carinena_2012}
J.~F. Cari\~{n}ena, M.~F. Ra\~{n}ada, M.~Santander, Curvature-dependent formalism, {Schrödinger} equation and energy levels for the harmonic oscillator on three-dimensional spherical and hyperbolic spaces, J. Phys. A 45~(26) (2012) 265303.
\newblock \href {https://doi.org/10.1088/1751-8113/45/26/265303} {\path{doi:10.1088/1751-8113/45/26/265303}}.

\bibitem{Redkov_2012}
V.~M. Red’kov, E.~M. Ovsiyuk, Quantum Mechanics in Spaces of Constant Curvature, Nova Science Publishers, New York, 2012.

\bibitem{Ozfidan_2025}
A.~\"{O}zfidan, Quantum pseudo-harmonic oscillator potential in non-{Euclidean} space: application to diatomic molecules, Phys. Scr. 100~(2) (2025) 025226.
\newblock \href {https://doi.org/10.1088/1402-4896/ada5cc} {\path{doi:10.1088/1402-4896/ada5cc}}.

\bibitem{Maclay_2020}
G.~J. Maclay, Dynamical symmetries of the {H} atom, one of the most important tools of modern physics: {SO(4)} to {SO(4,2)}, background, theory, and use in calculating radiative shifts, Symmetry 12~(8) (2020) 1323.
\newblock \href {https://doi.org/10.3390/sym12081323} {\path{doi:10.3390/sym12081323}}.

\bibitem{Infeld_1945}
L.~Infeld, A.~Schild, A note on the {Kepler} problem in a space of constant negative curvature, Phys. Rev. 67 (1945) 121--122.
\newblock \href {https://doi.org/10.1103/PhysRev.67.121} {\path{doi:10.1103/PhysRev.67.121}}.

\bibitem{Higgs_1979}
P.~W. Higgs, Dynamical symmetries in a spherical geometry. i, J. Phys. A 12~(3) (1979) 309--323.
\newblock \href {https://doi.org/10.1088/0305-4470/12/3/006} {\path{doi:10.1088/0305-4470/12/3/006}}.

\bibitem{Granovskii_1992}
Y.~I. Granovskii, A.~S. Zhedanov, I.~M. Lutsenko, Quadratic algebras and dynamics in curved spaces. {II}. the {Kepler} problem, Theor. Math. Phys. 91~(3) (1992) 604--612.

\bibitem{Sklyanin_1983}
E.~K. Sklyanin, Some algebraic structures connected with the {Yang-Baxter} equation, Funct. Anal. Its Appl. 16~(4) (1983) 263--270.

\bibitem{Carinena_2011}
J.~F. Cari\~{n}ena, M.~F. Ra\~{n}ada, M.~Santander, {The quantum free particle on spherical and hyperbolic spaces: A curvature dependent approach}, J. Math. Phys. 52 (2011) 072104.
\newblock \href {http://arxiv.org/abs/1201.5589} {\path{arXiv:1201.5589}}, \href {https://doi.org/10.1063/1.3610674} {\path{doi:10.1063/1.3610674}}.

\bibitem{With_2023}
G.~de~With, Melting is well-known, but is it also well-understood?, Chem Rev. 123~(23) (2023) 13713–13795.
\newblock \href {https://doi.org/10.1021/acs.chemrev.3c00489} {\path{doi:10.1021/acs.chemrev.3c00489}}.

\bibitem{Novikov_1984}
V.~Novikov, \href{http://jetp.ras.ru/cgi-bin/dn/e_060_03_0618.pdf}{Melting as a phase transition into a space with constant curvature}, JETP 60~(3) (1984) 618--623.
\newline\urlprefix\url{http://jetp.ras.ru/cgi-bin/dn/e_060_03_0618.pdf}

\bibitem{Kirchbach_2008}
M.~Kirchbach, C.~B. Compean, {Baryons from quarks in color gauge space of constant positive curvature and deconfinement} (2008).
\newblock \href {http://arxiv.org/abs/0805.2404} {\path{arXiv:0805.2404}}.

\bibitem{Quense_2023}
C.~Quesne, Quasi-exactly solvable extensions of the {Kepler–Coulomb} potential on the sphere, Ann. Phys. 451 (2023) 169265.
\newblock \href {https://doi.org/https://doi.org/10.1016/j.aop.2023.169265} {\path{doi:https://doi.org/10.1016/j.aop.2023.169265}}.

\bibitem{Carinena_2021}
J.~F. Cari\~{n}ena, M.~F. Ra\~{n}ada, M.~Santander, Superintegrability on the three-dimensional spaces with curvature. oscillator-related and {Kepler}-related systems on the sphere s3 and on the hyperbolic space h3, J. Phys. 54~(36) (2021) 365201.
\newblock \href {https://doi.org/10.1088/1751-8121/ac17a4} {\path{doi:10.1088/1751-8121/ac17a4}}.

\bibitem{Stevenson_1941}
A.~F. Stevenson, Note on the ``kepler problem" in a spherical space, and the factorization method of solving eigenvalue problems, Phys. Rev. 59 (1941) 842--843.
\newblock \href {https://doi.org/10.1103/PhysRev.59.842} {\path{doi:10.1103/PhysRev.59.842}}.

\bibitem{Nikiforov_1988}
A.~Nikiforov, V.~Uvarov, Special Functions of Mathematical Physics: A Unified Introduction with Applications, Springer, Basel, 1988.

\bibitem{Melnikov_1985}
V.~N. Mel'nikov, G.~N. Shikin, Hydrogen-like atom in the gravitational field of the universe, Sov. Phys. J. 28~(1) (1985) 47--51.
\newblock \href {https://doi.org/10.1007/BF00896051} {\path{doi:10.1007/BF00896051}}.

\bibitem{Ivashchuk_1996}
V.~D. Ivashchuk, V.~N. Melnikov, {Dually-charged mesoatom on the space of constant negative curvature}, J. Math. Phys. 37 (1996) 1642--149.
\newblock \href {http://arxiv.org/abs/math-ph/0504024} {\path{arXiv:math-ph/0504024}}.

\bibitem{Ellis_2023}
L.~Ellis, I.~Ellis, C.~Koutschan, S.~K. Suslov, {On Potentials Integrated by the {Nikiforov-Uvarov} Method} (2023).
\newblock \href {http://arxiv.org/abs/2303.02560} {\path{arXiv:2303.02560}}.

\bibitem{Tezcan_2009}
C.~Tezcan, R.~Sever, A general approach for the exact solution of the {Schr{\"o}dinger} equation, Int. J. Theor. Phys. 48~(2) (2009) 337--350.
\newblock \href {https://doi.org/10.1007/s10773-008-9806-y} {\path{doi:10.1007/s10773-008-9806-y}}.

\bibitem{Suslov_2020}
S.~K. Suslov, J.~M. Vega-Guzm{\'a}n, K.~Barley, An introduction to special functions with some applications to quantum mechanics, in: M.~Foupouagnigni, W.~Koepf (Eds.), Orthogonal Polynomials, Springer International Publishing, Cham, 2020, pp. 517--628.

\bibitem{Berkdemir_2012}
C.~Berkdemir, Application of the {Nikiforov-Uvarov} method in quantum mechanics, in: M.~R. Pahlavani (Ed.), Theoretical Concepts of Quantum Mechanics, IntechOpen, Rijeka, 2012, Ch.~11, pp. 225--252.
\newblock \href {https://doi.org/10.5772/33510} {\path{doi:10.5772/33510}}.

\bibitem{Zhang_2025}
P.~M. Zhang, Z.~K. Silagadze, P.~A. Horvathy, {Flyby-induced displacement: analytic solution} (2025).
\newblock \href {http://arxiv.org/abs/2502.01326} {\path{arXiv:2502.01326}}.

\bibitem{Kikuchi_2020}
I.~Kikuchi, A.~Kikuchi, Explanation of {Nikiforov-Uvarov} method in quantum mechanics – with a view toward algebraic geometry (2020).
\newblock \href {https://doi.org/10.31219/osf.io/4mzye} {\path{doi:10.31219/osf.io/4mzye}}.

\bibitem{Ismail_2005}
M.~Ismail, Classical and Quantum Orthogonal Polynomials in One Variable, Cambridge University Press, Cambridge, 2005.

\bibitem{Ballesteros_1993}
A.~Ballesteros, F.~J. Herranz, M.~A. del Olmo, M.~Santander, Quantum structure of the motion groups of the two-dimensional {Cayley-Klein} geometries, J. Phys. A 26~(21) (1993) 5801.
\newblock \href {https://doi.org/10.1088/0305-4470/26/21/019} {\path{doi:10.1088/0305-4470/26/21/019}}.

\bibitem{Wald_1984}
R.~M. Wald, {General relativity}, Chicago University Press, Chicago, 1984.

\bibitem{Raposo_2007}
A.~P. Raposo, H.~J. Weber, D.~E. Alvarez-Castillo, M.~Kirchbach, Romanovski polynomials in selected physics problems, Cent. Eur. J. Phys. 5~(3) (2007) 253--284.
\newblock \href {https://doi.org/10.2478/s11534-007-0018-5} {\path{doi:10.2478/s11534-007-0018-5}}.

\bibitem{Bessis_1979}
N.~Bessis, G.~Bessis, Electronic wavefunctions in a space of constant curvature, J. Phys. A 12~(11) (1979) 1991--1997.
\newblock \href {https://doi.org/10.1088/0305-4470/12/11/012} {\path{doi:10.1088/0305-4470/12/11/012}}.

\bibitem{Vinitsky_1993}
S.~I. Vinitsky, L.~G. Mardoian, G.~S. Pogosian, A.~N. Sisakian, T.~A. Strizh, {A Hydrogen atom in the curved space. Expansion over free solutions on the three-dimensional sphere}, Phys. Atom. Nucl. 56 (1993) 321--327.

\bibitem{Leemon_1978}
H.~I. Leemon, {Dynamical Symmetries in a Spherical Geometry. II}, J. Phys. A 12 (1979) 489--501.
\newblock \href {https://doi.org/10.1088/0305-4470/12/4/009} {\path{doi:10.1088/0305-4470/12/4/009}}.

\bibitem{Hadinger_1974}
G.~Hadinger, N.~Bessis, G.~Bessis, Closed‐form expressions of matrix elements and eigenfunctions from ladder‐operator considerations, J. Math. Phys. 15~(6) (1974) 716--726.
\newblock \href {https://doi.org/10.1063/1.1666716} {\path{doi:10.1063/1.1666716}}.

\bibitem{Duff_1949}
G.~F.~D. Duff, Factorization ladders and eigenfunctions, Can. J. Math. 1~(4) (1949) 379–396.
\newblock \href {https://doi.org/10.4153/CJM-1949-034-3} {\path{doi:10.4153/CJM-1949-034-3}}.

\bibitem{Arfken_2005}
G.~B. Arfken, H.~J. Weber, {Mathematical methods for physicists; 6th ed.}, Academic Press, San Diego, 2005.

\end{thebibliography}

\end{document}